\documentstyle[a4]{article}

\newcommand{\text}{\rm}
\newcommand{\be}{\begin{equation}}
\newcommand{\ee}{\end{equation}}
\newcommand{\bq}{\begin{eqnarray}}
\newcommand{\eq}{\end{eqnarray}}

\begin{document}

\title{{\bf  On the relation between the propagators of dual theories}}
\author{{\bf A.P. Ba\^{e}ta Scarpelli$^{a,b}$}, {\bf M. Botta Cantcheff$^{b}$}, \\
{\bf \thinspace and J.A. Helay\"el-Neto$^{a,b}\,$}
\thanks{{\tt E-mails:scarp@fisica.ufmg.br, botta.cbpf.br, 
helayel@cbpf.br.}}\vspace{2mm} \\
{\small {\bf $^{a}$}Grupo de F\'{\i}sica Te\'{o}rica  Jos\'{e} Leite Lopes}\\
{\small Petr\'{o}polis, RJ, Brazil.}\vspace{2mm}\\
{\small {\bf $^{b}$}CBPF, Centro Brasileiro de Pesquisas F\'{\i }sicas }\\
{\small Rua Xavier Sigaud 150, 22290-180 Urca} \\
{\small Rio de Janeiro, RJ, Brazil.}\vspace{2mm}.}
\maketitle

\begin{abstract}
\noindent

In this paper, we show that the propagator of the dual of a general Proca-like theory,
derived from the gauging iterative Noether Dualization Method, can be written by means
of a simple relation between known propagators. This result is also a 
demonstration that the Lagrangian obtained by dualization describes the same physical 
particles as the ones present in the original theory 
 at the expense of introducing 
new non-physical (ghosts) excitations.  
\end{abstract}

\vspace{0.5cm}

Recently, the so-called gauging iterative Noether Dualization Method (NDM) \cite{1}
has been
shown to be effective in establishing some dualities between models \cite{2}. This
method is based on the traditional idea of a local lifting of a global symmetry and
may be realized by 
an iterative embedding of Noether counterterms. However, this method provides  a strong
suggestion of duality since it has been shown to give the expected
result in the paradigmatic duality between the so-called Self-Dual
model and Maxwell-Chern-Simons in three dimensions (SD-MCS). This
well-known correspondence was first established in detail by Deser
and Jackiw \cite{3}, and may be shown by using a parent action approach \cite
{4}.

In a recent paper \cite{5} it was verified that, in calculating the dual of the
 modified 
eletromagnetic Maxwell Lagrangian, which includes Chern-Simons-like (that violates
 the Lorentz symmetry) and Proca terms, the corresponding propagator can
 be written 
as a simple relation with known propagators, namely:
\be
\langle A_\mu A_\nu \rangle= \langle A_\mu A_\nu \rangle _m-\langle A_\mu A_\nu
 \rangle _0 - 
\frac 1{m^2} \omega _{\mu \nu},
\label{1}
\ee
where $\langle A_\mu A_\nu \rangle _m$ corresponds to the propagator of the theory with the Proca term, 
$\langle A_\mu A_\nu \rangle _0$ is the propagator of the gauge invariant theory
 (massless) 
and $\omega _{\mu \nu}$ is the longitudinal spin operator $\omega _{\mu \nu}=
\partial_\mu 
\partial_\nu/\Box$.

This result was obtained for a specific theory \cite{5}. In this paper, we show that this is 
actually
a general result and besides that the gauging iterative Noether Dualization Method 
leads us to a Lagrangian that describes the same physical particles at the expense
of introducing non-physical 
excitations (ghosts). 

The method consists in introducing an auxiliary field, $B_\mu$, such that $\delta B_\mu= 
\delta A_\mu= \partial_\mu \eta$, in order to restore gauge invariance:
\be
{\cal L}_D={\cal L}-J_\mu B^\mu +\frac {m^2}{2}B_\mu B^\mu.
\ee
In the equation above, $\delta {\cal L}=J_\mu \delta A^\mu$, and $\delta J_\mu=m^2
 \delta A_\mu$,
by virtue of the Proca term. We should here stress that the latter relation is true 
whenever the Proca term is the only responsible for the breaking of the gauge symmetry.
So, the variation of the Lagrangian 
${\cal L}_D$ with respect to $B_\mu$ leads us to
\be
B_\mu=\frac 1{m^2} J_\mu,
\ee
and so
\be
{\cal L}_D={\cal L}-\frac 1{2m^2}J^2.
\ee

Let us then consider the two Lagrangians 
\be
{\cal L}=\frac 12 A{\cal O}A 
\ee
and
\be
{\cal L}_0=\frac 12 A{\cal O}_0A
\ee
obtained after suitable partial integrations in their respective actions. Here, we have omitted the Lorentz
 indices 
in order to simplify. In the equation above, ${\cal O}$ and ${\cal O}_0$ are
 differential (local) operators corresponding to the theories with and without the Proca
 term, respectively, and obey the relation
\be
{\cal O}_{\mu \nu}={\cal O}_{0\mu \nu} + m^2 g_{\mu \nu}.
\label{p1}
\ee
By applying now the NDM to the Lagrangian ${\cal L}$ we have
\be
\delta {\cal L}=\frac 12 \left \{ ({\cal O}A)\delta A + A({\cal O}\delta A)\right \}
= {\cal O}A \delta A
\label{cor}
\ee
The last step is done with the help of partial integration. The differential 
operator ${\cal O}$
 has second-order derivatives or, like in the topological case, a first-order 
 derivative contracted with the Levi-Civita
 tensor density. Equation (8) allows us to identify the Noether current as
\be
J={\cal O}A.
\ee
Now, we have to add to the Lagrangian
\be
-\frac 1{2m^2}J^2=-\frac 1{2m^2}({\cal O}A{\cal O}A)= -\frac 1{2m^2}A{\cal O}^2A,
\ee
where we again performed partial integrations. Our Lagrangian, then, becomes
\be
{\cal L}= \frac 12 A\left ( {\cal O}- \frac 1{m^2}{\cal O}^2 \right)A.
\ee
The complete wave operator can be finally identified as
\be
{\cal O}_T= {\cal O}- \frac 1{m^2}{\cal O}^2 = {\cal O}\left (
g- \frac 1{m^2}{\cal O} \right ).
\ee     
Using equation (7), we have
\be
g- \frac 1{m^2}{\cal O} =- \frac 1{m^2}{\cal O}_0 ,
\ee
so that 
\be
{\cal O}_T=- \frac 1{m^2}{\cal O}{\cal O}_0.
\ee

The propagator is defined as $\langle A_\mu A_\nu \rangle =
i\left( {\cal O}_T^{-1}\right )_{\mu \nu}$. It is simple to invert ${\cal O}_T$, once 
we know the inverses of ${\cal O}$ and ${\cal O}_0$. For the  ${\cal O}_0$-operator, it becomes necessary  
to add a gauge fixing term, $\frac {\Box}{\alpha} \omega_{\mu \nu}$. Therefore:
\be
{\cal O}_T=- \frac 1{m^2}{\cal O}\tilde {\cal O}_0,
\ee
with
\be
\tilde {\cal O}_0= {\cal O}_0 + \frac {\Box}{\alpha} \omega.
\ee
The inverse operator can then be readily written as
\be
{\cal O}_T^{-1}= -m^2 \tilde {\cal O}_0^{-1}{\cal O}^{-1}.
\ee

Now, we wish to show that this inverse operator is nothing but the relation presented in 
equation (1). In order to do this, we use that
\be
{\cal O}=\tilde {\cal O}_0 +m^2 g - \frac{\Box}{\alpha} \omega,
\ee
and then multiply both sides of the above equation at the right by ${\cal O}^{-1}$ 
and at the left by  $\tilde {\cal O}_0^{-1}$, so as to obtain
\be
\tilde{\cal O}_0^{-1}= {\cal O}^{-1} + m^2 \tilde {\cal O}_0^{-1}{\cal O}^{-1}
-\frac {\Box}{\alpha} \tilde {\cal O}_0^{-1} \omega {\cal O}^{-1}.
\label{inverso}
\ee
Now, the operators ${\cal O}^{-1}$ and $\tilde {\cal O}_0^{-1}$ can be split in the
form
\be
\tilde {\cal O}_0^{-1}=\tilde {\cal O}_{0GI}^{-1}+ \frac {\alpha}{\Box} \omega
\ee
and 
\be
{\cal O}^{-1}={\cal O}_{GI}^{-1}+ \frac {1}{m^2} \omega,
\ee
where the index GI is to indicate gauge invariance. This gauge invariance, in momentum 
space, is expressed by the relations below:
\be
k^\mu (\tilde {\cal O}_{0GI}^{-1})_{\mu \nu}=0
\ee
and
\be
k^\mu ({\cal O}_{GI}^{-1})_{\mu \nu}=0.
\ee 
So, since in momentum space the longitudinal spin operator is given by 
$\omega_{\mu \nu}=\frac {k_\mu k_\nu}{k^2}$, the last term in equation (19) 
will be read as
\be
-\frac {\Box}{\alpha} \left (\tilde {\cal O}_{0GI}^{-1}+ \frac {\alpha}{\Box} \omega
\right ) \omega \left ( {\cal O}_{GI}^{-1}+ \frac {1}{m^2} \omega \right )=
-\frac 1{m^2} \omega,
\ee
where we have used that $\omega ^2=\omega$. Finally, it is found that 
\be
-m^2 \tilde {\cal O}_0^{-1}{\cal O}^{-1}= {\cal O}^{-1}-\tilde {\cal O}_0^{-1}
-\frac 1{m^2} \omega,
\ee
or
\be
\langle A_\mu A_\nu \rangle= \langle A_\mu A_\nu \rangle _m-\langle A_\mu A_\nu
 \rangle _0 - 
\frac 1{m^2} \omega _{\mu \nu}. \nonumber
\ee

We then arrive at the decomposition proposed in eq. (1), which completes our demonstration.
The result above is general and is not necessarily to be 
restricted to the dimension D=4. An inspection of equation (1) shows 
that all the poles of the original theory, and therefore all the physical excitations, 
are still present in the new propagator. However, it happens that the poles 
of the massless theory, that originates the propagator $langle A_\mu A_\nu
\rangle _0$, behave as non-physical excitations, for this propagator is ruled by a minus 
sign, yielding negative residue at its poles, which correspond 
to negative-norm states or, in other words, ghosts. This final result also 
clarifies the real meaning of the duality implied by the Noether Dualization 
Method: the dual theories do not 
display the same spectrum; they share the same physical sector of their 
respective spectra. As already shown above, the dual theory carries 
non-physical modes due to the minus sign in front of $\langle A_\mu A_\nu \rangle _0$.

\section*{Acknowledgments}

The authors are indebted to CNPq and CLAF for the invaluable financial help. A. P. Ba\^eta Scarpelli ackowledges the Department of Physics of UFMG for the kind hospitality.

\end{document}